# Magnetization measurements on $Li_2Pd_3B$ superconductor

P. Badica[1)]

Institute for Materials Research, Tohoku University, 2-1-1 Katahira, Aoba-ku, Sendai, 980-8577, Japan and National Institute for Materials Physics, P.O. Box MG-7, 76900 Bucharest, Romania

T. Kondo, T. Kudo, Y. Nakamori, S. Orimo and K. Togano

Institute for Materials Research, Tohoku University, 2-1-1 Katahira, Aoba-ku, Sendai, 980-8577, Japan

Magnetization in DC magnetic fields and at different temperatures have been measured on the $Li_2Pd_3B$ compound. This material was recently found to show superconductivity at 7-8K. Critical fields $H_{c1}(0)$ and $H_{c2}(0)$ have been determined to be 135Oe and 4T, respectively. Critical current density, scaling of the pinning force within the Kramer model and the irreversibility field data are presented. Several superconductivity parameters were deduced: $\xi$=9.1 nm, $\lambda$=194nm and $\kappa$=21. The material resembles other boride superconductors from the investigated points of view.

[1)] Electronic address: p.badica@imr.tohoku.ac.jp

We have recently found superconductivity below 7-8K in the cubic $Li_2Pd_3B$ compound [1]. The material is a boride and the constituent elements are an alkaline element, Li and a late transition element, Pd. None of these elements were reported to be included in the binary or ternary boride superconductors as base ones, although alkaline earth, Mg, is a component of $MgB_2$ with surprisingly record-high $T_c$ of 39K and Pd belongs to the platinum group elements that are usually included in the boride superconductors (Ru, Rh, Os, Ir and Pt [2]). Also of interest is that Pd is a base element of the boride carbide compound Y-Pd-B-C showing the maximum record $T_c$ of 23K [3] in this class of superconductors.

In this letter we present the magnetization measurements performed by a SQUID magnetometer (Quantum Design 5T) with the purpose to extract the very first data on critical fields, critical current and superconductivity parameters. Evaluation of the $Li_2Pd_3B$ vs. other superconductors is important for generation of new directions in search of new superconductors targeting higher $T_c$, as well as for better understanding of superconductivity.

Material was synthesized by arc melting employing a two-step procedure: first ingots of $Pd_3B$ alloy were prepared and second, they were melted together with the Li-metal. The resulting material is $Li_2Pd_3B$ superconductor. Details on processing, $T_c$ and structure of the as-prepared superconductor are given elsewhere [1].

Magnetization loops M(H) taken at different temperatures are gathered in the Fig. 1, while the magnetization curves vs. temperature M(T), measured for several magnetic fields, are presented in the Fig. 2. Upper critical field ($H_{c2}^{M\text{-}H}$) was determined as the point where the M(H) loops are reaching the background (Fig. 1 inset (a)). Upper critical field ($H_{c2}^{M-T}$) was also estimated considering the onset point of the superconducting transition of the M(T) curves (Fig. 2). Variation of $H_{c2}^{M-H}$ and $H_{c2}^{M-T}$, as a function of temperature is shown in the Fig. 3. Linear fitting of the experimental points is resulting at T=0K in values of 5T and 5.2T, respectively (Fig. 3a). These values are comparable with 6.2T determined from measurements of electrical resistance in magnetic fields [1]. In the next calculations we have used the intermediate curve, i.e. for $H_{c2}^{M-T}$. In this case the absolute value of the initial slope $dH_{c2} / dT$ is 0.7T/K and is not far from the value, 0.8T/K for the Pd- containing boride-carbide Y-Pd-B-C [4]. Assuming the Werthamer-Hefland-Hohemberg (WHH) formula $H_{c2}(0)=0.691x(dH_{c2}/dT)_{Tc}x\ T_c$ [5] we obtain upper critical field $H_{c2}(0)$ of approximately 4T. It is easy to understand that lower $T_c$ than for some boride-carbides or $MgB_2$ is resulting in the lower value of the upper critical field. From the upper critical field ($H_{c2}=\Phi/2\pi\xi^2$, $\Phi$ is the flux quantum) we have extracted the coherence length $\xi$=9.1 nm. This is similar as for $MgB_2$ [2], it is in the upper limit for the boride-carbides [4] and it is considerably larger than for those of high $T_c$ cuprates (typically 2nm).

Another important parameter to characterize superconductivity is lower critical field $H_{c1}$. Values of $H_{c1}$ were taken as the field for which M(H) curve starts to deviate from the linear behavior generated by the perfect diamagnetism (Fig. 1 inset (b)). Data are plotted vs. temperature in the Fig. 3 inset. Fitting with $H_{c1}=H_{c1}(0)[1-(T/T_c)^2]$ is resulting in $H_{c1}(0)$=135Oe and $T_c$=7.5K. The value of $T_c$ from fitting agrees relatively well with the experimental value of 7.8K. The value of $H_{c1}(0)$ is considerably lower than for the boride-carbide superconductors (around 800Oe [4]) with high $T_c$ and is also approximately two-three times lower than for the $MgB_2$ [2], but is slightly higher than for the Re-boride superconductors (60-80Oe) [6]. From $H_{c1}(0)$ and $\xi$, the penetration depth can be calculated to be $\lambda$=194nm, by the formula $H_{c1}=(\Phi_0/4\pi\lambda^2)\ln(\lambda/\xi)$. This value is similar to the high limit of the $\lambda$-interval for the $MgB_2$ [2]. The Ginzburg-Landau parameter $\kappa=\lambda/\xi=21$. The extracted value is between the values reported for $Re_7B_3$ ($\lambda$=17nm) and $Re_3B$ ($\lambda$=35nm) superconductors [6].

Critical current density, $J_c$ vs. magnetic field evaluated by using Bean model [7] ($J_c=30\Delta M/d$, d is the diameter of the bulk sample in our case) is presented in the Fig. 4a. Values of $J_c$ are relatively low (e.g. $J_c$(0T, 2K)=5x10$^3$A/cm$^2$). The reason seems to be the presence of the cracks. To check this assumption we have measured M(H, 2K) loop on the part of the bulk sample after it was crushed into powder (Fig. 1 inset (c)). The resulted powder as well as the surface of the initial sample, showing cracks can be visualized in the Fig. 5. Curve of $J_c$ for the powder sample, considering the average grain size d=20μm in the Bean model, is plotted in the Fig. 4a. The value of $J_c$(0T, 2K) is 6x10$^4$A/cm$^2$ for the powder sample. Optical microscopy image on the polished cross section of the bulk sample [1] is indicating that the domains between cracks are approximately 5-10 times smaller than the size of the sample. Therefore, if in the Bean model, calculation is done by using the size of these domains, the values of $J_c$(0T) will be close to those observed for the powder sample. This suggests that the observed discrepancy in the values of $J_c$ is not due to the 'weak links' problem as for high $T_c$ superconductors.

Moreover, Kramer curves $J_c^{0.5}H^{0.25}(H)$ [8] from the Fig. 4b are linear in H and are similar to $Nb_3Sn$ [9] and $MgB_2$ [10], considered without weak links. Scaling with classic $h^{0.5}(1-h)^2$ function of the flux pinning force, $F_p=J_c\mu_0H$, is found within the whole measured interval of 2-7K (Fig. 4b inset). Extrapolation of the linear region of the Kramer curves results in $H_k(T)$-field values equivalent to empirical irreversibility line for which $J_c$ tends to zero. Data of $H_k(T)$ together with the experimental $H_{irr}(T)$ extracted from the M(H) loops for the same criterion of $J_c \rightarrow 0$ (Fig. 1 inset (a)) are presented in the Fig. 3. The $H_k(T)$ and $H_{irr}(T)$ lines are directly proportional to $H_{c2}(T)$. This result is similar to $MgB_2$ [10] and is very different from the high $T_c$ superconductors [11].

In summary, we have characterized Li-Pd-B compound using magnetization measurements. The material is shown to be a classic intermetallic boride superconductor from the investigated points of view. The material is expected to provide new directions in search for new superconductors with improved characteristics.

Authors thank E. Aoyagi and Y. Hayasaka of Tohoku University for help with SEM measurements and to K. Hirata of NIMS for useful discussions.

FIG. 1 Magnetization loops M(H) at 2, 3, 4, 5, 6 and 7K for the $Li_2Pd_3B$ bulk sample. Inset (a) and (b) are showing details of the same curves at high and low magnetic fields, respectively. The inset (c) shows M(H) loop at 2K for the sample in the powdered state.

FIG. 2 Magnetization curves vs. temperature for different applied magnetic fields. Inset is giving a detail showing the variation in the onset of the diamagnetic signal.

FIG. 3 Critical fields vs. temperature.

FIG. 4 Critical current density vs. applied magnetic field (a) for the $Li_2Pd_3B$ sample in the powder and bulk states and Kramer curves (b) for different temperatures of 2, 3, 4, 5, 6 and 7K (the curve for the highest temperature is closer to the left, low corner of the plot). The inset is presenting the scaling of the pinning force. Fitting with $h^{0.5}(1-h)^2$ is the dashed line.

FIG. 5 SEM (JEOL JSM 6700) images of the $Li_2Pd_3B$ sample in the (a) - powder and (b) - bulk state. On (b) image cracks are indicated by arrows.

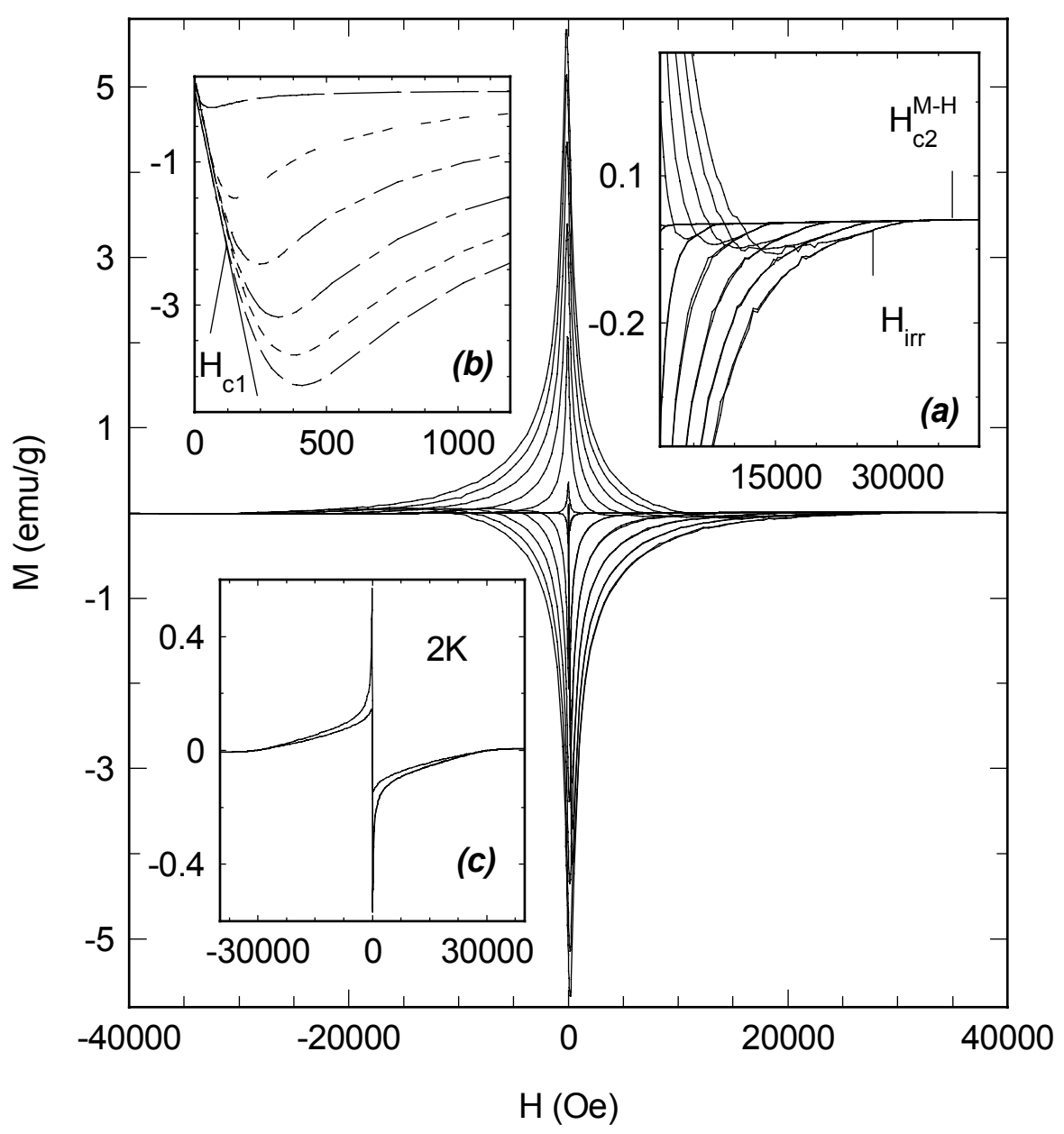

FIG. 1

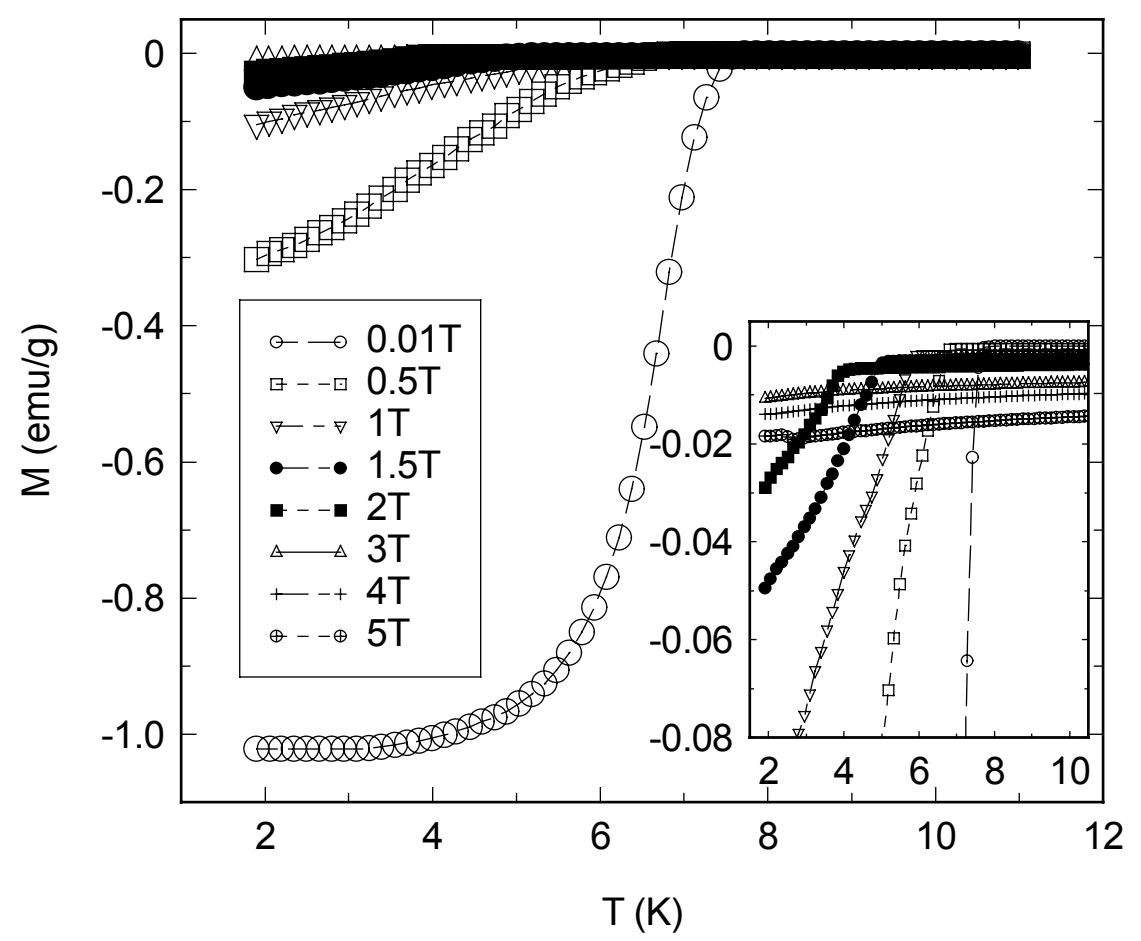

FIG. 2

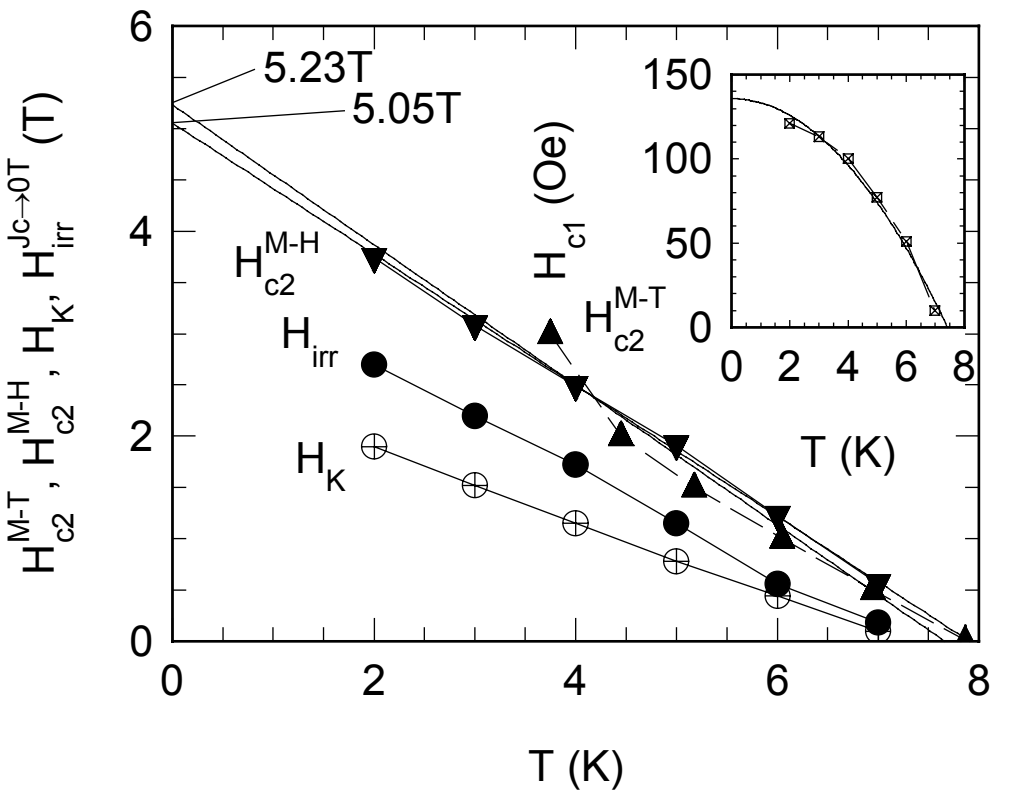
5.23T
5.05T
$H_{c2}^{M-H}$
$H_{irr}$
$H_K$
$H_{c1}$ (Oe)
$H_{c2}^{M-T}$
T (K)
$H_{c2}^{M-T}$, $H_{c2}^{M-H}$, $H_K$, $H_{irr}^{Jc\to 0T}$ (T)

FIG. 3

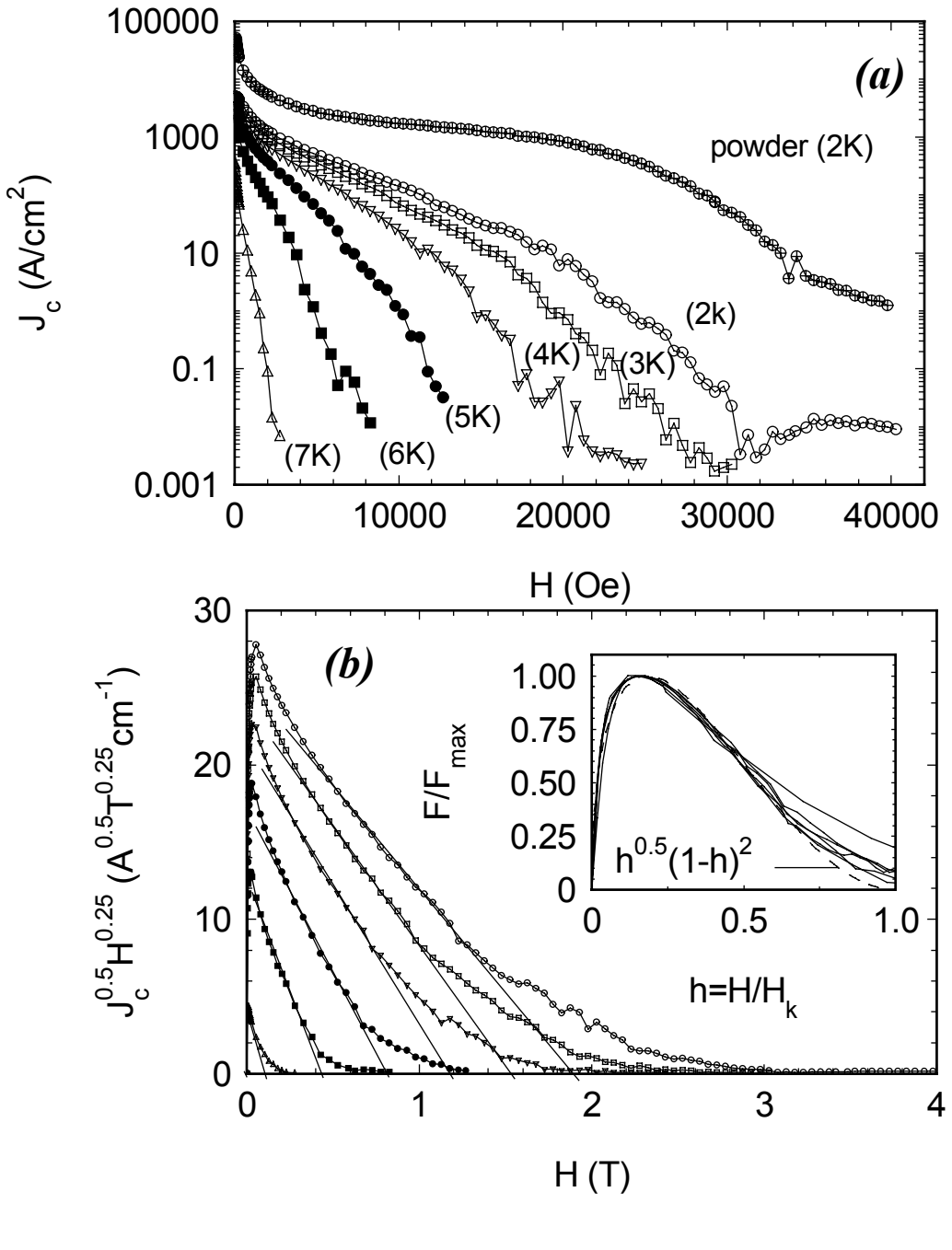

FIG. 4

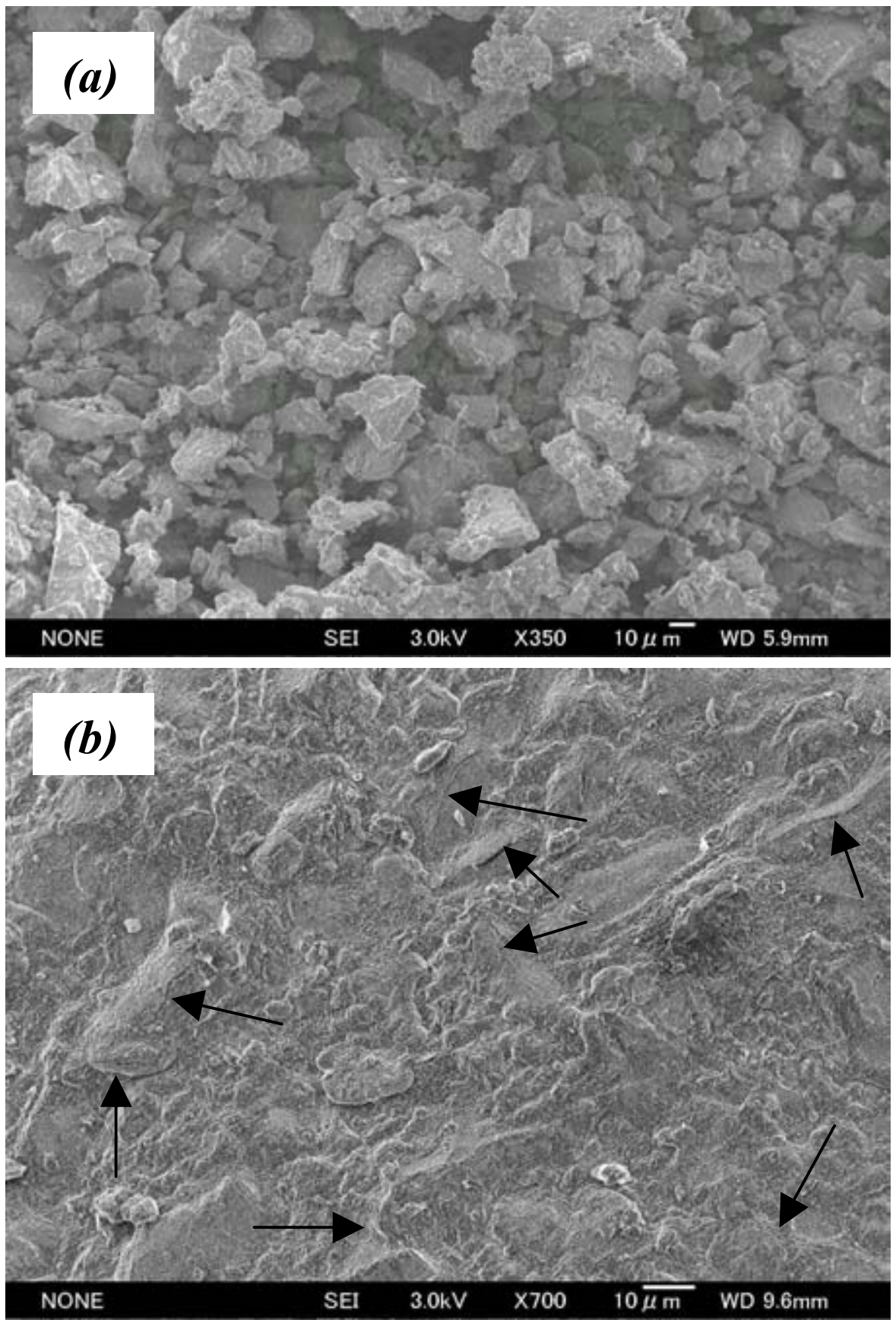
(a)
NONE SEI 3.0kV X350 10 μm WD 5.9mm
(b)
NONE SEI 3.0kV X700 10 μm WD 9.6mm

FIG. 5